\documentclass{aa}  

\usepackage{graphicx}
\usepackage{txfonts}
\usepackage{amsmath}

\newcommand{\TOriC}{\mbox{${\theta}^{1}$\, Ori C}}

\begin{document}

   \title{A triple-induced merger in \TOriC\, solves the proplyd lifetime problem and the curiously young shell of M42}

   \titlerunning{A triple-induced merger solution to the ONC's lifetime problems}

   \author{M. J. C. Wilhelm\inst{1}
          \and
           A. A. Trani\inst{2,3}
          \and
           S. Portegies Zwart\inst{1}
          }

   \institute{Leiden Observatory, Leiden University, PO Box 9513, 2300 RA, Leiden, The Netherlands
         \and
         Department of Astronomy, University of Concepci\'on, Avenida Esteban Iturra s/n
         Casilla 160-C Concepci\'on, Chile
         \and
         National Institute for Nuclear Physics – INFN, Sezione di Trieste, I-34127, Trieste, Italy
         }
   \date{Received ...; accepted ...}

 
  \abstract
   {There has long been a tension between the stellar ages in the Orion Nebula Cluster and the rapid external photoevaporation of its protoplanetary discs. Recently, the shell of the Orion Nebula has similarly been found to be dynamically younger than the stars.}
   {We propose a new scenario to resolve both age problems. In our scenario the ONC's most massive star, \TOriC1, formed from the merger between two binary stars induced by vZLK oscillations due to the present-day binary companion \TOriC2. This merger, happening between the last episode of star formation and today, resulted in an increased luminosity, leading to stronger external photoevaporation and accelerated expansion of the shell.}
   {We use Monte Carlo simulations to explore both whether a sudden brightening of \TOriC1 could reconcile the stellar ages and shell expansion, and whether triple progenitors to \TOriC\, could result in stellar mergers on the relevant timescale.}
   {A sudden brightening of \TOriC1 a few hundred thousand years ago can reproduce today's shell while maintaining a system age close to the stellar age. However, the required increase in stellar wind luminosity is greater than a simple merger can explain. At lower luminosity, the mixing of a relatively greater amount of cool photoevaporative wind material into the hot bubble could result in the required increase, and we identify a lower limit on the required strength of this effect. Additionally, we find that progenitor triples merge readily, on timescales even shorter than the scenario would require. These mergers could be deferred by increased pre-main sequence tidal effects, rotation, and perturbing encounters with other cluster members.}
   {Our proposed scenario with a merger driven by von
     Zeipel-Lidov-Kozai evolution solves two problems; the Orion
     Nebula's proplyd problem and the shell lifetime problems.}


   \maketitle
 %

\section{Introduction}

Early Hubble Space Telescope observations of the Orion Nebula revealed various dark, comet-like tailed structures \citep{Odell1994}. These were quickly identified to be protoplanetary disks with winds driven by the radiation of nearby massive stars, in particular \TOriC1 \citep{Henney1998,Henney1999}, and were dubbed proplyds (with the process of mass loss due to radiation from nearby massive stars being dubbed external photoevaporation). A problem quickly arose, as these proplyds were losing mass at rates up to $\sim$10$^{-6}$ M$_\odot$ yr$^{-1}$, so that at the inferred age of the Orion Nebula Cluster (ONC) of $\sim$1 Myr \citep{Hillenbrand1998}, their initial masses should have been comparable to the masses of their host stars. This runs counter to observations of lower masses for young protoplanetary disks \citep{Tychoniec2018} and our theoretical understanding of the gravitational stability of circumstellar disks \citep{Haworth2020}.

More recently, the expansion velocity of the shell component of the Orion Nebula was measured to be $\sim$13 km/s by \citet{Pabst2019,Pabst2020}, which combined with a radius of $\sim$2.5 pc, results in an age of $\sim$0.3 Myr. The expansion of this shell is primarily driven by the stellar winds of the ONC's massive stars. It is striking that two processes powered by the ONC's massive stars both appear to have younger dynamical ages than the inferred stellar age.

One solution to the proplyd lifetime problem was offered by
\citet{Winter2019}. Their scenario is based on observations by
\citet{Beccari2017} arguing for populations of three different ages in
the ONC, ranging from nearly 3 to about 1 Myr old \citep[see also][for
  slightly younger inferred ages; they also propose a physical origin
  of this multimodality in the repeated dynamical ejection of massive
  stars that disrupt star formation, leading to multiple periods of
  collapse and expansion of the natal gas cloud]{Kroupa2018}. The
model of \citet{Winter2019} posits that each population of stars forms
in a subvirial state, with highly elliptical orbits around the cluster
centre. This brings them close to where massive stars end up due to
dynamical mass segregation, and subsequent virialization sets them on
orbits that reach further from the cluster centre \citep[this is
  corroborated by the observations of][which show that the younger
  population is also more centrally concentrated]{Beccari2017}. This
naturally results in a flow of young, less irradiated discs to the
regions of high irradiation. They also reproduce the weak correlation
between the masses of the discs and their host stars (in contrast to
other regions, and as expected from massive stars having deeper
gravity wells that allow them to retain their discs for longer).

\citet{Winter2019} are also able to roughly reproduce the present-day fraction of stars with discs while retaining high mass loss rates due to external photoevaporation. However, their simulations only include stars of at least 0.3 M$_\odot$. In typical stellar initial functions, roughly half of all stars are less massive than that, and their discs will have shorter lifetimes due to less massive initial discs and shallower gravity wells. While they show that a lower disc viscosity results in a larger disc fraction (less viscous spreading results in discs remaining deeper in the stars' gravity wells), this is done by decreasing the overall external photoevaporation rate, contradicting observations. An additional factor may thus be needed to explain the large disc fraction when including low-mass stars.

The three population scenario has recently been challenged by \citet{Alzate2023}. They propose that the three populations that \citet{Beccari2017} interpreted as having different ages, may well be single stars, unresolved binaries, and unresolved triples. They also find that star formation extended up to 0.3 Myr ago, which would entirely solve the proplyd and shell lifetime problems. However, their sample extends beyond the ONC, and the sample of \citet{Beccari2017} has greater completeness and more accurate photometry in the core of the ONC, which may have led them to mis the footprint of the populations of different ages. The proplyd and shell lifetime problems thus remain unsolved.

Here, we introduce a new scenario that can resolve the proplyd
lifetime problem and the shell lifetime problems with a single
mechanism. We propose that the most massive star of the ONC, \TOriC1,
brightened considerably sometime between today and the end of the most
recent period of star formation. This brightening may have occurred
through a merger between two stars resulting in the single star
currently known as \TOriC1. This merger was then induced by von
Zeipel-Lidov-Kozai (vZLK) oscillations force by the present-day binary
companion, \TOriC2. Because a star's energy production scales more
strongly than linear with mass, a single star will be more luminous
than two stars of the same total mass. This brightening accelerated
both external photoevaporation and the expansion of the wind-driven
shell; extrapolating the present-day rates backwards in time then
naturally results in underesting its age.

Our scenario is fully compatible with that of \citet{Winter2019};
effectively, we propose that one massive cluster member\footnote{The
binary \TOriC\, is sufficiently tight that it can plausibly be
approximated as a single star in simulations of this type.} of the
youngest population increased in luminosity at an intermediate point
in time. The period where the radiation field was lower would have
contributed to the survivability of discs around low mass stars.

Scenarios that solve the proplyd lifetime problem that rely on a
recent event are often critiqued using a statistical argument: we are
unlikely to find the ONC in a short-lived configuration. However, we
argue that this is mostly an issue for events that should have
happened on the order of $10^4$ yr ago; a timescale of a few $10^5$
yrs ago corresponds to a few tens of percents of the time since the
ONC's formation, implying a considerable probability of finding the
system in the aftermath of such an event; which scales roughly
linearly with time.

A merger can explain two other properties of \TOriC1: the oblique
magnetic field observed by \citet{Stahl1996}, and the small nitrogen
excess observed by \citet{Rzaev2021}. An oblique merger, where the
spin axes are misaligned with the orbital axis, can set up
differential and misaligned rotation in the merger product, which then
effectively and efficiently generates a magnetic field
\citep{Ferrario2009,Schneider2019}. The merger product would inherit
most of the initial stars' angular momentum, but magnetic braking due
to stellar winds can subsequently slow rotation \citep{Mestel1968} to
the relatively slow rotation at the present day
\citep{SimonDiaz2006,SimonDiaz2007}.

A merger of two massive stars would also mix material from the core, which is burning hydrogen in the CNO cycle, into the envelope of the merger product. This is suspected to be the source of the nitrogen excess in the blue straggler $\tau$ Sco \citep{Martins2012,Keszthelyi2021}. \citep{Martins2012}, however, was unable to confirm a nitrogen excess in \TOriC1\, due to its young stellar age, \citet{Rzaev2021} later did find a small excess compared to other early type stars in the ONC. While smaller than that of $\tau$ Sco, it was found from the combined spectrum of \TOriC1\, and \TOriC2, and additional observations are needed to confirm or disprove a nitrogen excess in \TOriC1.

This paper is structured as follows. In Sections \ref{sec:expansion} and \ref{sec:triple}, we present our two lines of investigation. In Section \ref{sec:expansion} we compare different models of shell expansion to the various observed properties of the Orion Nebula and its stellar cluster. In Section \ref{sec:triple} we explore the properties and evolution of the progenitor triple and stellar merger. In both sections, we include subsections discussing the underlying methods (\ref{subsec:expansion_methods} and \ref{subsec:triple_methods}), describing the specific models used (\ref{subsec:expansion_models} and \ref{subsec:triple_models}), reporting the results (\ref{subsec:expansion_results} and \ref{subsec:triple_results}), and discussing our intermediate findings (\ref{subsec:expansion_discussion} and \ref{subsec:triple_discussion}). Then, in Section \ref{sec:conclusion}, we present our final conclusions.

\section{Shell expansion models} \label{sec:expansion}

\subsection{Methods} \label{subsec:expansion_methods}

The problem of the expansion of a stellar wind-driven bubble into the surrounding interstellar medium (ISM) allows for a simple analytic solutions in the form of a power law combining the stellar wind luminosity, ISM density, and time. \citet{Weaver1977} presented the solution for a uniform ISM density and a constant wind luminosity. This model was expanded to non-uniform ISM density distributions by \citet{Ostriker1988} and bubbles covering less than a full sphere by \citet{Geen2022}. 

The basic equations of motion of the bubble radius $R_W$ consist of momentum and energy balance:

\begin{equation}
\begin{split}
\quad& \frac{d}{dt} \left(M\left(R_W\right)\frac{dR_W}{dt}\right) = \frac{2E_b}{R_W}, \\
\quad & \frac{d}{dt} E_b = L_W - \frac{2E_b}{R_W}\frac{dR_W}{dt}.
\end{split}  
\end{equation}
Here $M\left(R_W\right)$ is the ISM mass initially contained within $R_W$ (and now swept up in the bubble wall), $E_b$ is the energy contained within the bubble, and $L_W$ is the stellar wind luminosity. 

Following \citet{Geen2022}, we define the ISM to be a decreasing power law of radius from the wind source ($\rho\sim r^{-\omega}$), and that the stellar wind only couples into $\Omega$ steradians of surrounding material (corresponding to the assumption that the ISM in the remaining $4\pi-\Omega$ steradians is dense enough to not be affected by the stellar wind), leading to the following definition of $M\left(R_W\right)$:

\begin{equation}
\quad M\left(R_W\right) = \frac{\Omega}{3-\omega} \rho_0r^3\left(\frac{r_0}{r}\right)^\omega.
\end{equation}
Here $\rho_0$ and $r_0$ define the density distribution. Without loss of generality, we fix $r_0$ to 1 pc, thus defining $\rho_0$ to be the ISM density at 1 pc. 

Assuming that the bubble does not lose energy through radiative losses, its energy turns out to be:

\begin{equation}
\quad E_b = \frac{5-\omega}{11-\omega}L_Wt.
\end{equation}

Assuming a constant $L_W$, these equations have the following solution:

\begin{equation}
\quad R_W\left(t\right) = \left(A_W\left(\omega,\Omega\right) \frac{L_Wt^3}{\rho_0r_0^\omega}\right)^{1/\left(5-\omega\right)}.
\end{equation}
Here $A_W$ is a numerical factor of order unity, defined by \citet{Geen2022}. Note that setting $\omega=0$, corresponding to a uniform ISM, indeed reduces to the solution of \citet{Weaver1977} again. We can now also easily acquire the bubble expansion velocity $V_W$ using differentiation:

\begin{equation}
\quad V_W\left(t\right) = \frac{3}{5-\omega}\left(A_W\left(\omega,\Omega\right) \frac{L_W}{\rho_0r_0^\omega t^{2-\omega}}\right)^{1/\left(5-\omega\right)} = \frac{3}{5-\omega}\frac{R_w}{t}.
\end{equation}

This model can also accommodate a gradual brightening of the wind luminosity by defining it as a power law as well, which would yield a similar analytic solution. However, our triple merger model is characterized by a rapid increase in $L_W$ better modelled as a step function, which is not straightforwardly solved analytically.

Instead, we can model an instantaneous brightening by taking the analytic solution for the pre-brightening luminosity, at the moment of brightening, and using that as initial conditions for a numerical integration of the differential equations using the post-brightening luminosity. We can use a similar technique to explore another proposed solution of the proplyd lifetime problem, namely that the bubble has previously been confined to a dense core and has only recently `broken free'. This scenario can be modelled as a discontinuity in the density distribution rather than the wind luminosity.

The equations and parameters of this problem are few enough, and the timescale short enough, that we can statistically explore it. Using Markov Chain Monte Carlo (MCMC), we can introduce observation knowledge (with uncertainties) as priors and attempt to constrain unknown parameters such as pre-brightening luminosity and the moment of brightening. This also gives us a statistical framework to compare the different scenarios.

\subsection{Models} \label{subsec:expansion_models}

We use the {\sc emcee} package \citet{2013PASP..125..306F} to perform the MCMC, with 32 walkers and consider the chain as converged after $50l$ iterations, where $l$ is the longest autocorrelation length among the model parameters, and discard as burn-in the first $2l$ iterations.

For numerical integration, we use 4$^\textrm{th}$ order Runge-Kutta integration, with a time-step equal to $10^{-2}\textrm{min}\left(|y/\frac{dy}{dt}|\right)$, where $y$ is a vector containing all integrated quantities, and the absolute value is taken element-wise.

We use the shell's observed radius of $2.5\pm 0.125$ pc and observed
expansion velocity of $13\pm 2$ km/s as measurements, and the system
age and other model variables as free parameters in the model. In the
following we define which parameters each model has, and which priors
we use for each.

We explore three families of models. The {\it classical} family uses the analytic solution and serves as a null experiment. This model has 4 parameters: the system age (or present time), $t_1$, the wind luminosity, $L_{W}$, the shell mass $M_s$, and the density power law index $\omega$. Note that the shell mass is uniquely related to a density $\rho_0$ for a given $t_1$, $R_W$, and $\omega$. 

The {\it brightening} family uses numerical integration with a varying wind luminosity, $L_{W,0}$ before and $L_{W,1}$ after brightening, which happens at a time $\Delta t$ before $t_1$. $t_1$, $M_s$, and $\omega$ are the same as in the classical models, so with 2 new parameters we end up with a total of 6.

The {\it top hat} family also uses numerical integration with two different ISM reference densities at 1 pc, which is $\rho_0$ within a core radius $r_c$ and $\rho_1$ outside of that. $t_1$, $L_W$, and $\omega$ remain as in the classical model, again adding 2 new parameters for a total of 6\footnote{We have kept $\omega$ the same before and after the density jump as a total of 6 parameters allows easy comparison to the brightening models.}.

We follow \citet{Geen2022} in introducing two further varieties of
each model. \TOriC\, is actually located near the edge of the Orion
Nebula's shell, and the back wall appears to not be expanding, which
implies that expansion is not spherical, but rather covering about a
quarter of a sphere. Therefore, we perform our analysis for two values
of $\Omega$, namely $\pi$ and $4\pi$, which we refer to as the {\it
  quarter} and {\it full} sphere models. For the quarter sphere
models, the shell is actually the full bubble diameter away from
\TOriC, and therefore we use a shell radius of $5\pm 0.25$ pc for
those models (while the expansion velocity remains unchanged).

For the prior on the system age $t_1$, we take a normal distribution in $\log_{10}\left(t/\textrm{Myr}\right)$ with a mean of 0.0086 and a standard deviation of 0.2 (such that $t_1=1.02_{-0.38}^{+0.60}$
Myr). This corresponds to the mean of the ages for the youngest population in the ONC from the works of \citet{Beccari2017,Kroupa2018} with a larger standard deviation to reflect the uncertainties in the stellar age. For the time since the merger $\Delta t$, we take a flat distribution between 0 and $t_1$.

For the present-day wind luminosity of \TOriC1\, $L_{W,1}$, we use a normally distributed prior in $\log_{10}\left(L_{W,1}/\left(\text{erg/s}\right)\right)$ with a mean of 35.8 and a standard deviation of 1. In this we follow \citet{Pabst2019} (and references therein), but with a larger standard deviation. For $L_{W,0}$, the wind luminosity before brightening, we take a flat distribution in $\log_{10}\left(L_{W,0}/\left(\text{erg/s}\right)\right)$ between $\log_{10}\left(L_{W,1}/\text{(erg/s)}\right)-4$ and $\log_{10}\left(L_{W,1}/\left(\text{erg/s}\right)\right)$. 

For the cloud density $\rho_0$, we take as prior a distribution in $\log_{10}\left(\rho_0/\left(\text{g/cm}^{-3}\right)\right)$ such that the total shell mass $M_s$ is normally distributed in $\log_{10}\left(M_s/M_\odot\right)$, with a mean of $\log_{10}\left(1500\right)$ and a standard deviation of 0.2. This approximately corresponds, at $2\sigma$, to the lower and upper limits on the shell mass \citep{Pabst2019,Pabst2020}. For the density jump in the top hat models, we use as prior for $\log_{10}\rho_1/\rho_0$ a flat distribution between -9 and 0.

For the density gradient $\omega$ we adopt a flat distribution between
0 (a uniform distribution) and 3 (while keeping both, mass and radius,
finite) as prior. For the core radius $r_c$ we take a flat
distribution in
$\log_{10}\left(r_c/R_W\left(t_1;\rho_0,L_{W,0},\omega\right)\right)$
between -6 and 0; the core radius is then between one and one
millionth of a classical bubble's extent, but with a preference for
larger radii.

We summarize all of these priors in Table \ref{tab:shell_priors}.

\begin{table*}
\begin{tabular}{|l|l|l|l|}
Model(s) & Quantity & Prior & References \\
\hline
All & $\log_{10} t_1/\text{Myr}$ & $\mathcal{N}\left(\log_{10} \left(1.02\right), 0.2\right)$ & \cite{Beccari2017,Kroupa2018} \\
& $\log_{10} L_{W,1}/\left(\text{erg/s}\right)$ & $\mathcal{N}\left(35.8,1\right)$ & \cite{Pabst2019} \\
& $\log_{10} M_s/\text{M}_\odot$ & $\mathcal{N}\left(\log_{10}\left(1500\right),0.2\right)$ & \cite{Pabst2019,Pabst2020} \\
& $\omega$ & $\mathcal{U}\left(0,3\right)$ & \\
\hline
Brightening & $\Delta t$ & $\mathcal{U}\left(0,1\right)\times t_1$ & \\
& $\log_{10} L_{W,0}$ & $\mathcal{U}\left(-4,0\right) + \log_{10} L_{W,1}$ & \\
\hline
Top hat & $\log_{10}\left(\rho_1\right)$ & $\mathcal{U}\left(-9, 0\right) + \log_{10}\rho_0$ & \\
& $\log_{10}\left(r_c\right)$ & $\mathcal{U}\left(-6, 0\right) + \log_{10}\left(R_W\left(t_1;\rho_0,L_{W,0},\omega\right)\right)$ & 
\end{tabular}
\caption{The priors on the accelerated expansion shell model used in the MCMC.}
\label{tab:shell_priors}
\end{table*}

\subsection{Results} \label{subsec:expansion_results}

\begin{figure*}
    \centering
    \includegraphics[width=0.66\linewidth]{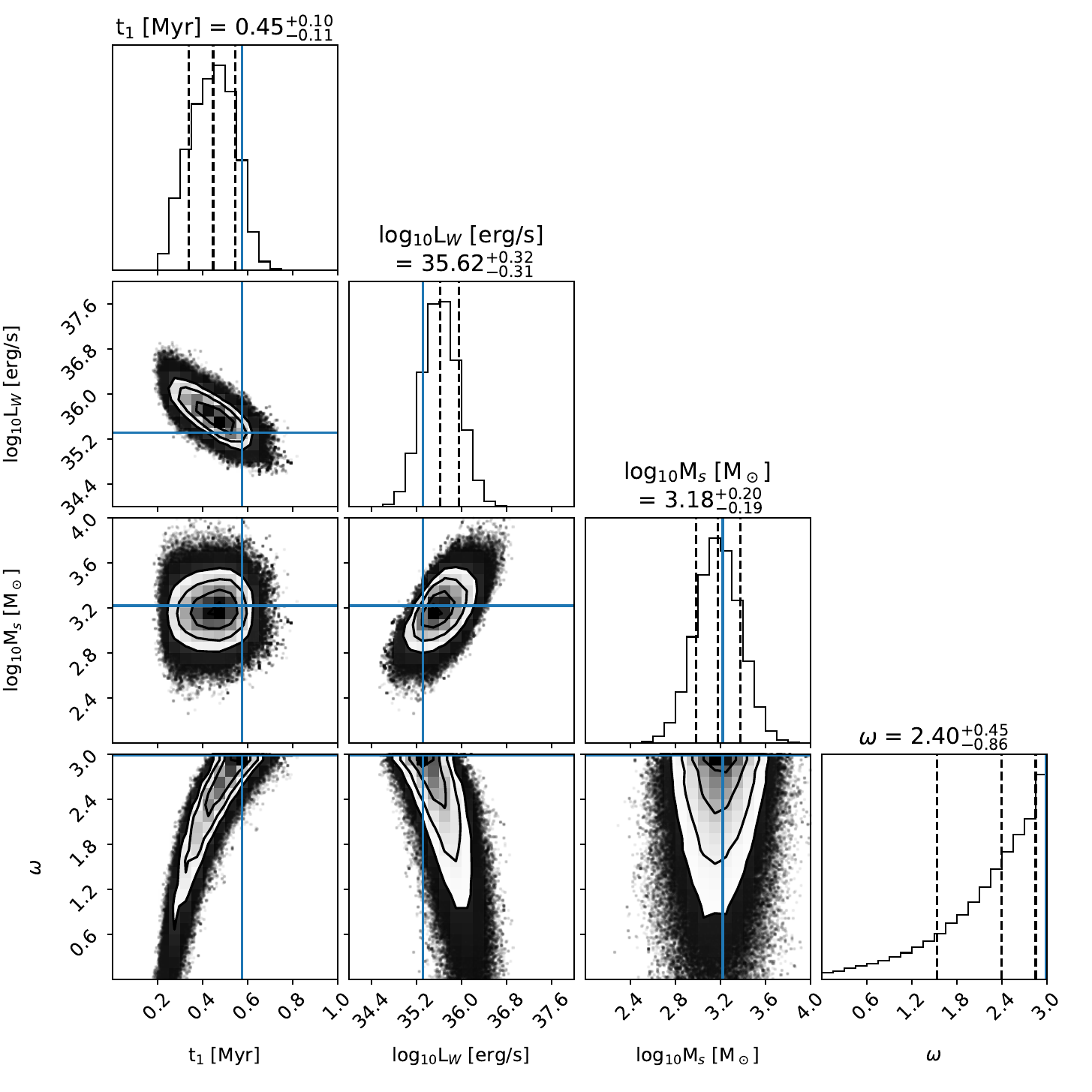}
    \caption{A corner plot of the posterior distribution of the classical quarter sphere model. Diagonal panels show one-dimensional probability distributions of each parameter, and off-diagonal panels show the two-dimensional probability distributions of the parameters in the same row and column. Blue lines indicate the parameters of the maximum-likelihood sample found by the MCMC. \\The parameters shown are the system age $t_1$, stellar wind luminosity $L_W$, the shell mass $M_s$, and the density power law $\omega$.}
    \label{fig:corner_classical_quarter}
\end{figure*}

We ran 6 MCMCs models, each with 4 or 6 free parameters. These result
in a lot of corner plots, most of which we present in the
Appendix~\ref{sec:appendix1} to preserve the flow of the paper. Here
we present the corner plots for the 3 quarter sphere configuration
models. In fig.\,\ref{fig:corner_classical_quarter}, we present the
quarter sphere model, which leads to a larger present-day bubble, and
the model satisfactoritly explains the shell lifetime.

In Figure \ref{fig:corner_classical_quarter} we show the posterior
distribution of the classical quarter sphere model. Notably, the
posteriors on the wind luminosity and shell mass remain consistent
with their priors, but the system age's posterior mean has shifted 
towards $\log_{10}\left(0.45\right)\approx-0.35$, nearly 2$\sigma$
from its prior, and the posterior probability of a system age of
$\approx$1 Myr is approaching 0. The model also favours a centrally
concentrated initial gas distribution, with a strictly increasing,
non-Gaussian posterior for the density gradient $\omega$.

\begin{figure*}
    \centering
    \includegraphics[width=\linewidth]{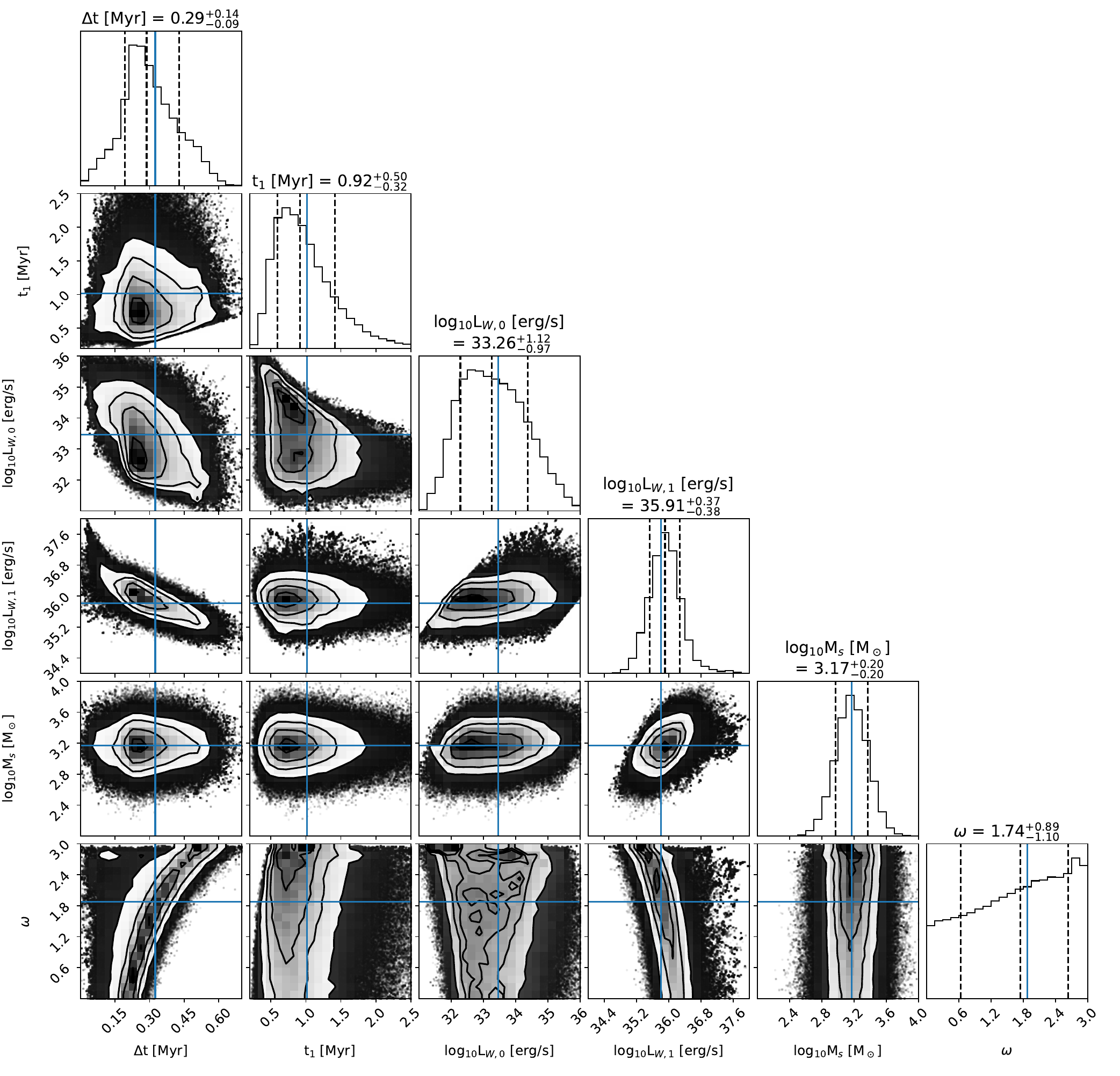}
    \caption{A corner plot of the posterior distribution of the brightening quarter sphere model. Diagonal panels show one-dimensional probability distributions of each parameter, and off-diagonal panels show the two-dimensional probability distributions of the parameters in the same row and column. Blue lines indicate the parameters of the maximum-likelihood sample found by the MCMC. \\The parameters shown are the time since brightening $\Delta t$, the system age $t_1$, the stellar wind luminosity before ($L_{W,0}$) and after ($L_{W,1}$) brightening, the shell mass $M_s$, and the density power law $\omega$.}
    \label{fig:corner_brightening_quarter}
\end{figure*}

In Figure \ref{fig:corner_brightening_quarter} we show the posterior distribution of the brightening quarter sphere model. As in the classical model, the posteriors of the present-day wind luminosity and shell mass remain consistent with their priors. This time, the posterior of the system age also remains broadly consistent with its prior. 

The brightening model is not as strongly constrained by its priors as the classic model, but the three parameters with flat priors have posteriors that yield some amount of information. The posteriors of the time since brightening and initial wind luminosity, while not properly Gaussian in shape, are centrally peaked with wings moving towards zero. However, the correlation between the initial and final wind luminosities reveals that there is a maximum initial wind luminosity, below which any value is as good another. The posterior of the density gradient does not move towards zero anywhere on its prior domain, but mostly increases towards larger values. 

These posterior distributions imply that while the brightening model as a whole is not constrained, we can constrain the time since brightening and put an upper limit on the initial wind luminosity. The time since brightening is typically on the order of a few $10^5$ years, and the wind luminosity must have increased by a factor of at most $\sim$500. These parameters allow system ages of $\sim$1 Myr. 

\begin{figure*}
    \centering
    \includegraphics[width=0.49\linewidth]{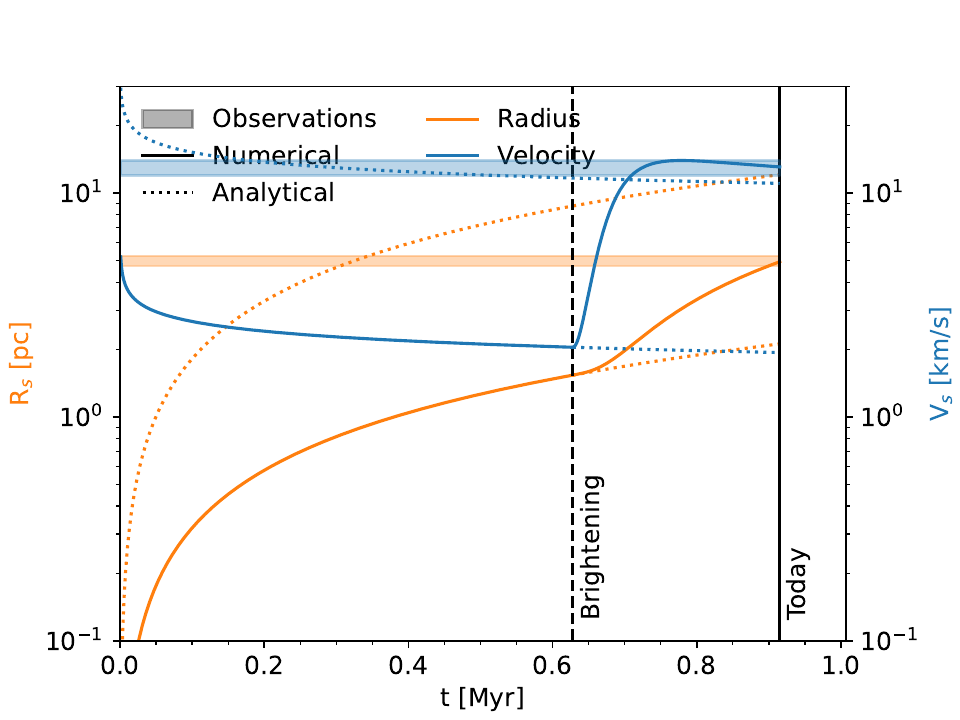}\includegraphics[width=0.49\linewidth]{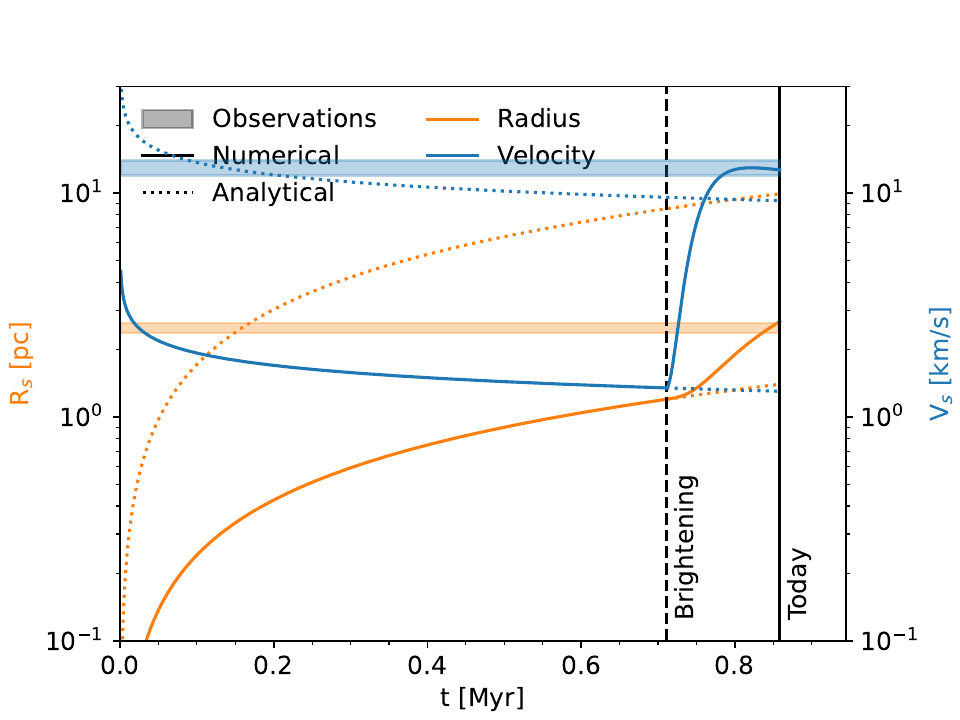}
    \caption{The two maximum probability solutions of the brightening model for the quarter (left) and full (right) sphere models. Solid lines show the models themselves, while the dotted lines show the analytical solutions for the initial and final wind luminosity. The horizontal shaded regions show the observed shell velocity and (respectively) the bubble diameter and radius.}
    \label{fig:expansion_model_brightening}
\end{figure*}

In Figure \ref{fig:expansion_model_brightening} we show the maximum-likelihood solutions found by the MCMC (parameters corresponding to the blue lines in Figure \ref{fig:corner_brightening_quarter}) for both the quarter and full sphere brightening models. Both are consistent with the shell's present-day size and expansion velocity. In addition to the numerical solutions we also show the analytic solutions for both the initial and final wind luminosities. At the moment of brightening, the numerical solutions accelerate away from the analytic solutions for the initial wind velocity, reaching a mostly constant expansion speed after about 0.1 Myr. This velocity is higher than the analytic solution for the final wind luminosity, allowing the shell's size to eventually catch up to this analytic solution. 

\begin{figure*}
    \centering
    \includegraphics[width=\linewidth]{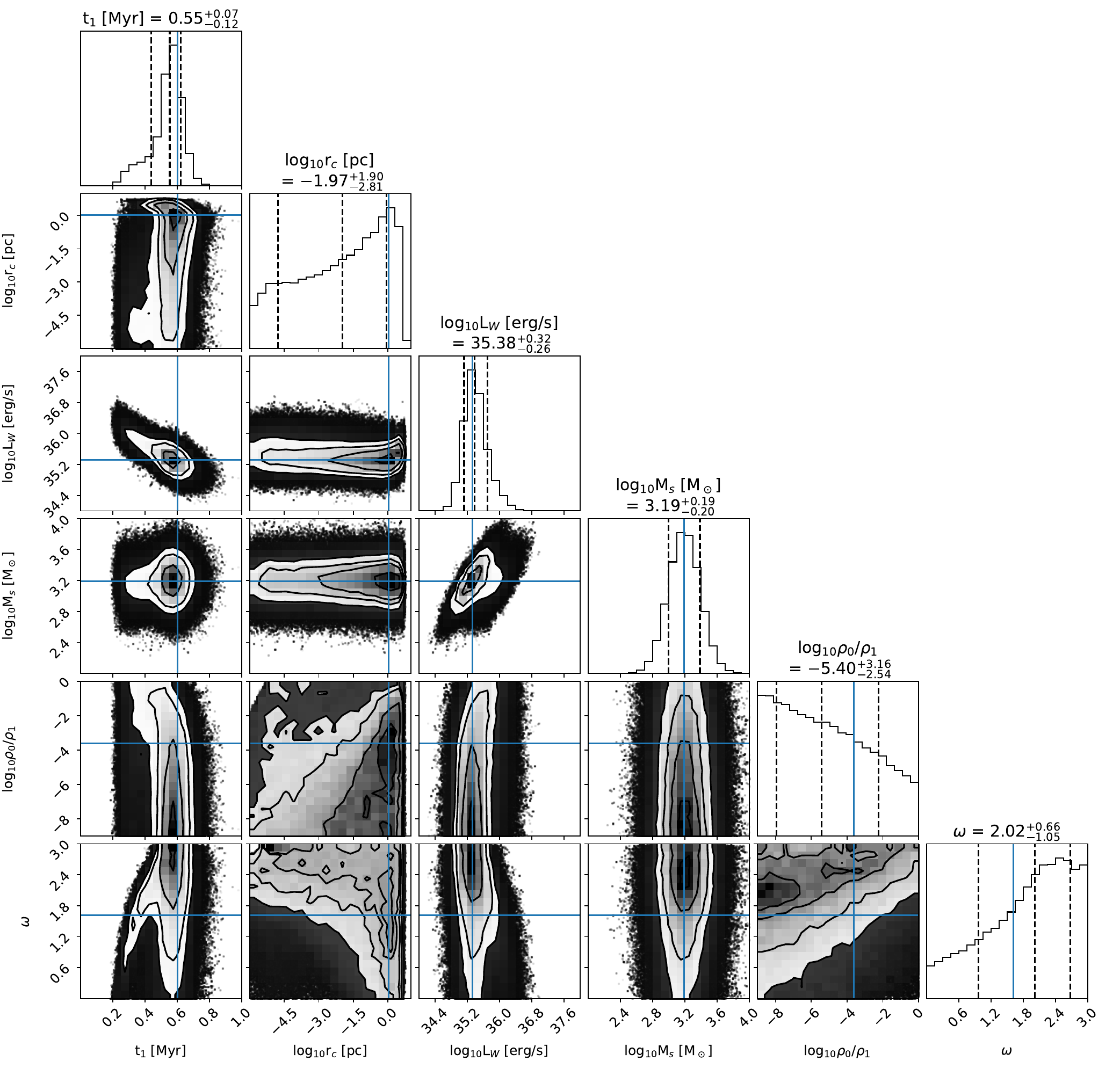}
    \caption{A corner plot of the posterior distribution of the top hat quarter sphere model. Diagonal panels show one-dimensional probability distributions of each parameter, and off-diagonal panels show the two-dimensional probability distributions of the parameters in the same row and column. \\The parameters shown are the system age $t_1$, the core radius $r_c$, the stellar wind luminosity $L_{W}$, the shell mass $M_s$, the density contrast between the core and beyond $\rho_0/\rho_1$, and the density power law $\omega$.}
    \label{fig:corner_tophat_quarter}
\end{figure*}

In Figure \ref{fig:corner_tophat_quarter} we show the posterior distribution of the top hat quarter sphere model. Here again, the posterior distributions of the wind luminosity and the shell mass remain consistent with the priors. However, the core radius, density ratio and density gradient remain largely unconstrained, respectively typically favouring larger, smaller, and larger values. Meanwhile, the system age is constrained rather strongly, but to values only slightly larger than in the classic model. Notably, the probability density approaches 0 above 0.8 Myr. 

\begin{figure*}
    \centering
    \includegraphics[width=0.49\linewidth]{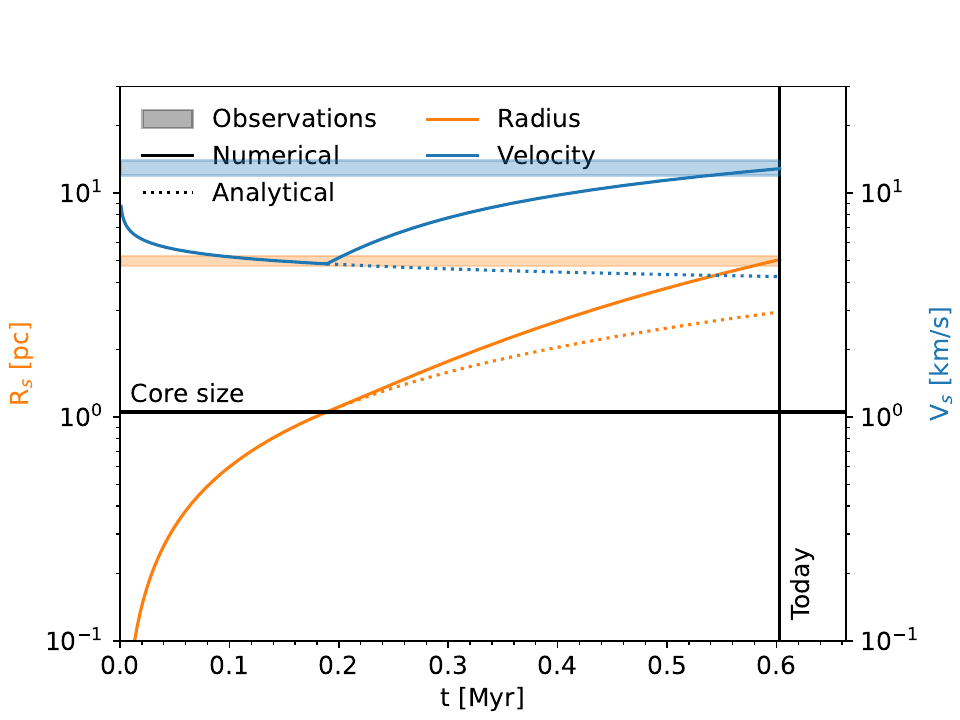}\includegraphics[width=0.49\linewidth]{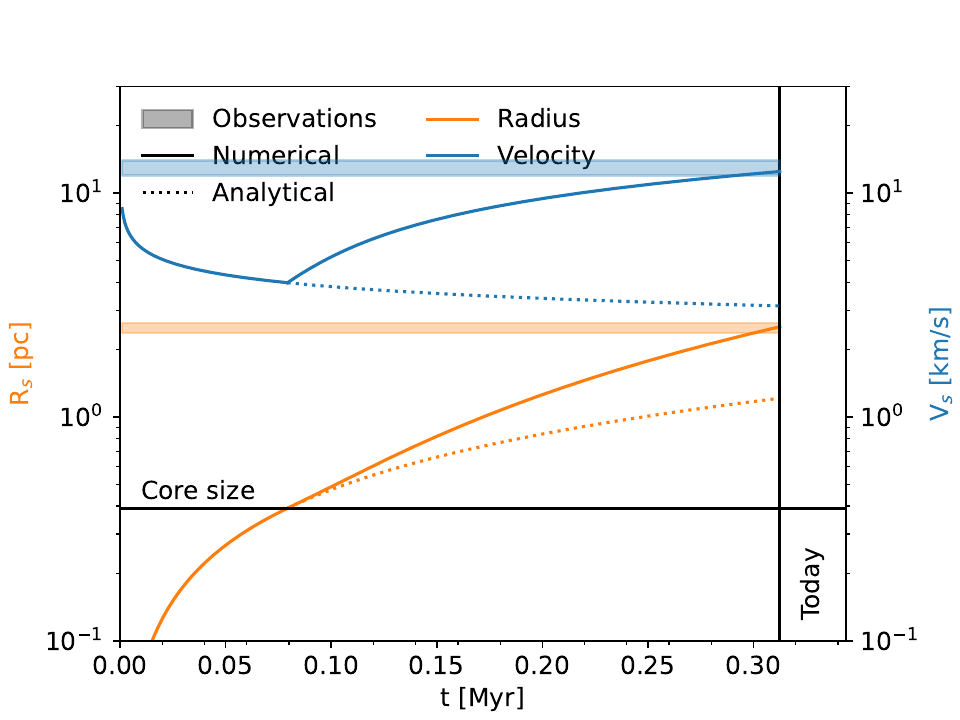}
    \caption{The two maximum probability solutions of the top hat model for the quarter (left) and full (right) sphere models. Solid lines show the models themselves, while the dotted lines show the analytical solutions for the core's density. The horizontal shaded regions show the observed shell velocity and (respectively) the bubble diameter and radius.}
    \label{fig:expansion_model_tophat}
\end{figure*}

In Figure \ref{fig:expansion_model_tophat} we show the maximum-likelihood solutions found by the MCMC for both the quarter and full sphere top hat models. Both are consistent with the shell's present-day size and expansion velocity. In addition to the numerical solutions we also show the analytic solutions for the core's density. At the moment the shell grows to the core radius, the numerical solutions accelerate away from the analytic solution. Unlike the brightening model, this acceleration is more gradual, and continues to the present day.

\subsection{Discussion} \label{subsec:expansion_discussion}

\subsubsection{Model comparison}

\begin{table}
    \centering
    \begin{tabular}{|l|l|l|}
         & \multicolumn{2}{l|}{Log-likelihood} \\
        Model & Quarter sphere & Full sphere \\
        \hline
        Classical & -0.967 & -3.859 \\
        Brightening & {\bf -0.005} & {\bf -0.007} \\
        Top hat & -0.795 & -3.486 \\
    \end{tabular}
    \caption{The maximum log-likelihood of the samples of all six MCMC
      runs. The two greatest log-likelihoods is marked in bold.}
    \label{tab:likelihoods}
\end{table}

In Table \ref{tab:likelihoods} we list the maximum log-likelihoods of
all six models. Comparing the classical model with the other models in
this way is not entirely fair, because the increase in free parameters
allows for more freedom in fitting to the data. The Akaike and
Bayesian information criteria (respectively AIC and BIC) can correct
for this, introducing a term that penalizes for the number of
parameters $k$, but these assume a large sample size $n$, at the very
least much larger than $k$. However, in our framework, we have a
sample size of 2 measurements (shell radius and expansion velocity),
and 4 or 6 (not entirely, due to informative priors) free
parameters. Tellingly, the AIC has a correction factor for small
sample sizes that is {\it negative} when $k\ge n$, thus favouring {\it
  more} model parameters. Still, we can properly compare the
brightening with the top hat models.

Overall, the quarter sphere models fit the constraints better than the full sphere models. Interestingly, this trend is strongest for the classical and top hat models (with both quarter sphere models also fitting better than each other's full sphere model), but the brightening model fits better than the other model families for both sphere geometries, and the quarter sphere model fits only slightly better than the full sphere model. 

Additionally, it is striking that the brightening models improve the log-likelihood much more than the top hat models, even tough they add the same number of parameters.

Finally, it's worth noting that for the classical and top hat models, the probability density of $t_1$ nearly vanishes above 0.4 Myr for the full sphere models, and above 0.8 Myr for the quarter sphere models, meaning they are still in tension with the inferred cluster age of the ONC.

\subsubsection{Required luminosity increase}

\begin{figure}
    \centering
    \includegraphics[width=\linewidth]{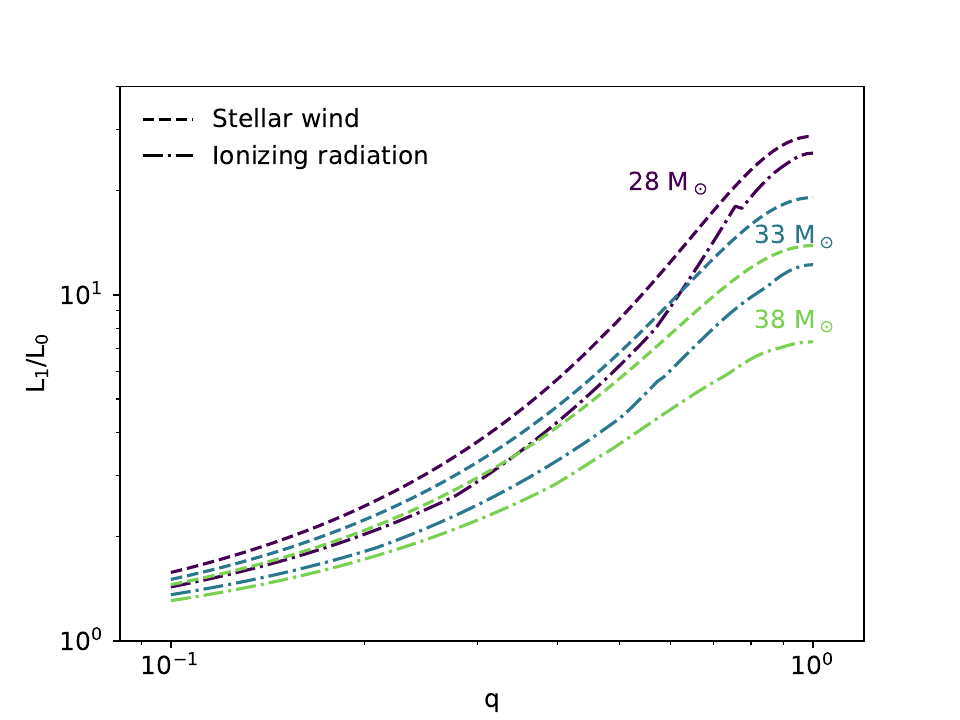}
    \caption{The luminosity ratio of a single star of given mass to that of two stars of equal total mass, as a function of the two stars' mass ratio $q$. Dashed lines show the stellar wind luminosity ratio, while the dash-dotted lines show the ionizing radiation luminosity ratio. Colors indicate the total mass (with lighter colors for more massive stars).}
    \label{fig:luminosity_ratio}
\end{figure}

The correlation between the wind luminosity before and after
brightening implies that \TOriC\, should have brightened by at least a
factor $\sim$10$^2$. However, a fully conservative merger of two stars
with a total mass equal to that of \TOriC\, would increase the stellar
wind luminosity by at most $\sim$10$^{1.5}$, as demonstrated in Figure
\ref{fig:luminosity_ratio}; this increase is maximized for stars of
comparable mass, a lower total mass, and a more
conservative merger.

However, \citet{Arthur2012} showed that the expansion of a stellar wind bubble can be slowed by the mixing of low-velocity photoevaporative winds into the hot gas of the bubble (a process we'll refer to as proplyd mass loading). The radial expansion of these proplyd mass loaded bubbles behaves like non-mass loaded bubbles, up to a constant scaling factor; from the data points in their Figure 6, we extract power law indices of the proplyd mass loaded radius evolution of 0.61 (panel a) and 0.55 (panel b), close to the analytic value of $3/5$. The constant scaling factor varied per model, but was up to $\sim$2. Translating this efficiency factor to an effective wind luminosity (such that $L_W\rightarrow\eta L_W$), this results in a factor $1/\eta\sim 2^5\approx 10^{1.5}$.

Proplyd mass loading is effective before and after the merger, but its efficiency depends on the ratio of stellar wind mass loss to proplyd mass loss, $\dot{M}_W/\dot{M}_p$. $\dot{M}_p$, in turn, depends on the ionizing luminosity $L_\mathrm{EUV}$ with $\dot{M}_p\sim L_\mathrm{EUV}^{1/2}$ \cite{Johnstone1998}. Because the ionizing luminosity increases by a similar amount as (even slightly less than) the stellar wind luminosity upon merger, the efficiency factor will increase with the merger. \cite{Arthur2012} did not explore the behaviour of this efficiency factor, but the accelerated expansion scenario requires an additional luminosity increase of at least $10^{0.5}$. Assuming that the efficiency $\eta$ scales as $\eta\sim\left(\dot{M}_W/L_\mathrm{EUV}^{1/2}\right)^\alpha$, and that $\dot{M}_W/L_\mathrm{EUV}^{1/2}$ increases by a factor $\sim$10$^{0.5}$\footnote{Both $L_W$ and $L_\mathrm{EUV}$ increase by $\sim$10, and the stellar wind velocity is relatively flat compared to the stellar mass loss rate.} then $\alpha$ must be at least 1.

\subsubsection{Implications for the proplyd lifetime problem}

An increase in the ionizing luminosity on merger of $\sim$10$^1$ would
result in an increase in photoevaporation rate of
$\sim$10$^{0.5}\approx 3$. This in turn would decrease the disc
lifetime by about the same amount. Combined with the dynamical effects
of \citet{Winter2019}'s scenario, this helps resolve the proplyd
lifetime problem.

The slower initial expansion does little to help shield the protoplanetary discs from \TOriC's radiation field. The shell radius reaches 0.3 pc within 0.1 Myr for the brightening models (Figure \ref{fig:expansion_model_brightening}) and within 0.05 Myr for the top hat models (Figure \ref{fig:expansion_model_tophat}), leaving proplyds in the direct neighbourhood of \TOriC\, exposed to its ionizing radiation for most of their lifetime.

\section{Triple models} \label{sec:triple}

\subsection{Methods} \label{subsec:triple_methods}

Simulating the long-term evolution of three stars of (broadly)
comparable mass is challenging due to their chaotic equations of
motion. Hierarchical triples typically have stable orbits over many
orbits, but following many cycles of the vZLK process still requires
careful integration. Additionally, non-Newtonian effects are
significant on the scales we're working in: the extended nature of
stars introduces tidal effects, and the high system density introduces
general relativistic effects. Both of these effects can dampen the
vZLK effect \citep{Naoz2014}.

We adopt the direct N-body code {\sc tsunami} \citep{Trani2023} to integrate the star's
equations of motion, called through the Astronomical Multipurpose
Software Environment (or {\sc amuse}, for short)
\cite{2009NewA...14..369P,2018araa.book.....P,2026araa.book.....P}
{\sc tsunami} was developed specifically for systems with a small
number of gravitating bodies, and includes modules for general
relativity \citep[to 3.5th order,][]{Blanchet2014}, secular dynamical
stellar tides \citep{Zahn1975,Hut1981,Eggleton1998}, and rotation,
which we all use. Additionally, {\sc tsunami} allows collisions as
stopping conditions.

We use a Monte Carlo approach to explore the likelihood of collisions occurring in a vZLK-susceptible triple where the outer binary resembles the present-day orbit of \TOriC\, and the inner binary has a total mass equal to \TOriC1. By sampling the full parameter space allowed by the present-day observational constraints we can assess the feasibility of a progenitor configuration for the merger scenario, and of this merger happening on a timescale within the required timescale. This timescale is about 1 Myr, so we continue integration until that time unless a merger occurs or the triple becomes unbound.

\subsection{Models} \label{subsec:triple_models}

We use the stars' physical radii as collisional radii (although
disruptions also likely happen when stars approach at distances
slightly larger than these radii). We obtain these radii (along with
the gyration radii, used in the tidal model later on) from the {\sc
  SeBa} parametrized stellar evolution code
\citep{PortegiesZwart1996,Toonen2012} through the {\sc amuse} framework, at
the ZAMS and solar metallicity. The apsidal motion constants are
instead derived from {\sc mse} \citep{Hamers2021} \citep[which were in
  turn derived from the stellar models of][]{Claret2004}. We use
polytropic indices of 1.5, and calculate the time lag as in
\citet{Hurley2002}. Because this time lag depends on the orbital
parameters, we recompute it with a frequency of ten times per vZLK
timescale as derived by \citet{Kiseleva1998}.

We compute spin periods using physical stellar radii and rotational
velocities. We randomly select the latter from the fit to the distribution of
rotational velocities of \citet{Garmany2015}, who cover early B stars
and the very latest O stars. To correct for the projection factor
$\sin i$ we divide the scale parameters by $2/\pi$, which is the mean
of $\sin\left(\mathcal{U}\left(0,\pi/2\right)\right)$. The spin axes
are initially aligned with the orbital axis of the inner binary.

The initial conditions of the outer binary are based on the present-day properties of \TOriC\, as measured by \citet{Balega2015}. For the outer binary, we uniformly cover the parameter space that is susceptible to vZLK oscillations. However, to combat the curse of dimensionality we aim to reduce degeneracies as much as possible, such as rotations of the same configuration and different phases of the same vZLK oscillation. We fix the orientation of the outer binary's orbit, and start from initially circular inner binaries that have an inclination within the critical Kozai angles \citep{Naoz2014}.

We randomly select the total mass of the inner binary from a normal distribution with a mean of 33.5 M$_\odot$ and a standard deviation of 5.2 M$_\odot$, with a mass ratio $q$ selected from a uniform distribution between 0.5 and 1. The mass of \TOriC2 is selected from a normal distribution with a mean of 12 M$_\odot$ and a standard deviation of 3 M$_\odot$.

We randomly select the semimajor axis $a$ of the inner binary from a log-uniform deviation between 0.3 au and the stability limit \citep{Vynatheya2022}, and that of the outer binary such that the outer period is normally distributed with a mean of 11.28 yr and a standard deviation of 0.02 yr.

We randomly select the relative inclination of the inner and outer binary $i_{rel}$ from a uniform distribution between $\cos^{-1}\sqrt{3/5}$ and $\pi-\cos^{-1}\sqrt{3/5}$ (the critical Kozai angles), and fix the outer binary's orbital plane to be the $x-y$ plane.  The inner binary is taken to be initially circular, and the eccentricity of the outer binary is taken from a normal distribution with a mean of 0.59 and a standard deviation of 0.01.  The argument of periapsis $\omega_p = 0$ for both the inner and outer orbit, because the inner orbit is circular and the outer orbit is in the equatorial plane.  For the outer orbit the longitude of the ascending node $\Omega = 0$, and for the inner orbit it is taken from a uniform distribution in the range $[0,2\pi]$.  The mean anomalies $M$ of the inner and outer orbit are randomly selected from uniform distributions in the range $[0,2\pi]$.  

We clipped the normally distributed stellar masses to within $\pm 1\sigma$ of the mean as these are upper and lower limits more than Gaussian errors. We also clip the other normally distributed parameters to $\pm 3\sigma$ to prevent extreme outliers. The full initial conditions are summarized in Table \ref{tab:lk_ics}.

\begin{table*}
\centering
\begin{tabular}{|l|l|l|l|}
 & Inner binary & Outer binary & References \\
\hline
$M$ [M$_\odot$] & $\mathcal{N}\left(33.5,5.2\right)$ & $\mathcal{N}\left(12,3\right)$ & \cite{Balega2015} \\
$q$ & $\mathcal{U}\left(0.5,1\right)$ & N/A & \\
$\log_{10} a/\text{au}$ & $\mathcal{U}\left(\log_{10} 0.3,\{\textrm{Stability}\}\right)$ & (from period) & \cite{Vynatheya2022} \\
$P$ [yr] & (from semimajor axis) & $\mathcal{N}\left(11.28,0.02\right)$ & \cite{Balega2015} \\
$e$ & 0 & $\mathcal{N}\left(0.59,0.01\right)$ & \cite{Balega2015} \\
$i$ [rad] & \multicolumn{2}{l|}{$\mathcal{U}\left(\cos^{-1}\sqrt{3/5},\pi-\cos^{-1}\sqrt{3/5}\right)$} & \\
$\Omega$ [rad] & $\mathcal{U}\left(0,2\pi\right)$ & 0 & \\
$\omega$ [rad] & 0 & 0 & \\
$M$ [rad] & $\mathcal{U}\left(0,2\pi\right)$ & $\mathcal{U}\left(0,2\pi\right)$ & \\
$P_{spin}$ & \multicolumn{2}{l|}{See text} & \\
$\hat{n}_{spin}$ & \multicolumn{2}{l|}{Aligned with inner binary} & \\
\end{tabular}
\caption{The initial conditions of the Monte Carlo simulations of the progenitor triple system of \TOriC. Note that the mass of the inner binary is the total mass, with the mass ratio chosen separately, while that of the outer binary is that of the tertiary. The inclination is relative between the inner and outer binary. }
\label{tab:lk_ics}
\end{table*}

\subsection{Results} \label{subsec:triple_results}

\begin{figure*}
    \centering
    \includegraphics{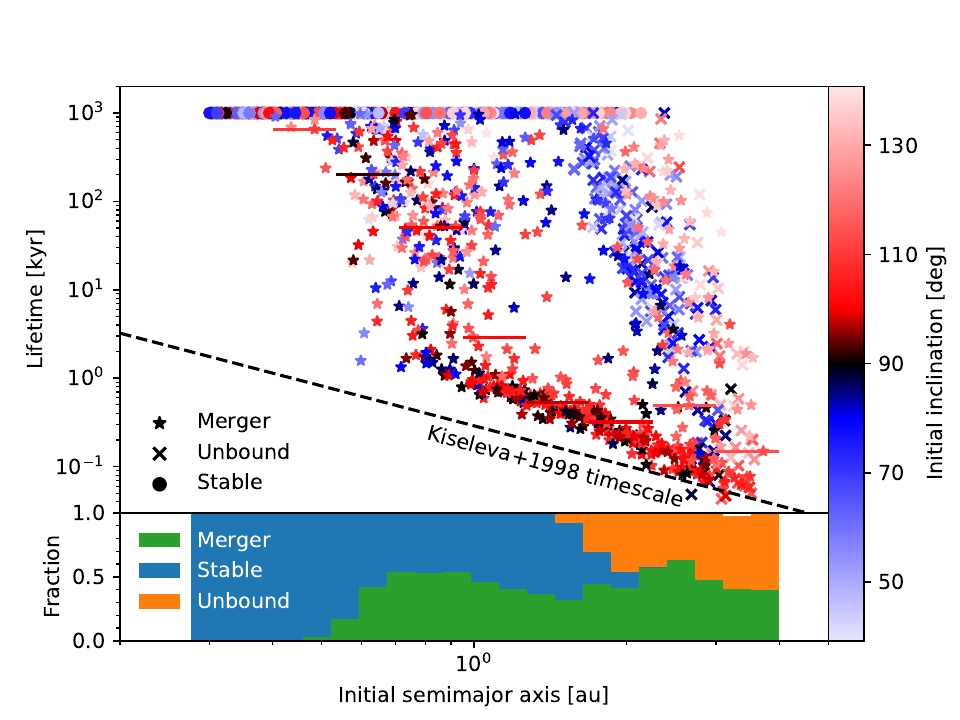}
    \caption{The fate of the simulated progenitor triples of \TOriC. The outer binary is the same as the present-day binary, while the inner binary has a semimajor axis as shown on the on the $x$-axis, is circular and vZLK-unstable, and has a total mass equal to \TOriC. {\it Top:} the lifetime of 1992 triples as a function of the initial semimajor axis of the inner binary. The outcome states are merger (star), unbound (cross), or stable for at least 1 Myr (point). The initial relative  inclination between the inner  and the outer orbit of the triples are indicated with colors (scale to the right).  The dashed curve indicates the vZLK timescale \cite{Kiseleva1998}. The horizontal dashes indicate median merger times for the  indicated range in orbital separation, and are colored by the median initial inclination of all merged triples in that range. {\sc Bottom:} The fraction of outcome states per range of initial semimajor axis. }
    \label{fig:triple-fate}
\end{figure*}

Figure \ref{fig:triple-fate} shows the fate of the 1992 simulated triples; both whether they ended up merging, unbound, or stable for at least 1 Myr, and when a merger or unbinding happened. A number of interesting clusters can be seen. Many merger events cluster along a line similar to the vZLK-timescale of \cite{Kiseleva1998}, though slightly steeper. These merger events mostly originate from initially retrograde orbits, and happen quickly, within hundreds or thousands of years. A very steep and slightly broader band containing both mergers and unbound systems can be seen at semimajor axes of $\sim$1.5 au and beyond. Finally, there is a cluster of merger events at intermediate semimajor axes ($\sim$0.5-1 au) spread over a large range of system lifetimes. Interestingly, in an earlier series of runs that did not include rotation, this region was barely populated. Finally, within $\sim$0.5 au, the number of merger events decreases sharply. This can be explained by tidal and/or relativistic effects circularizing the inner orbit. 

Outside of this circularization region, roughly half of all systems merge within 1 Myr. From just within 2 au and outward, the remaining non-merging triples instead become dynamically unstable. 

\begin{figure}
    \centering
    \includegraphics[width=\linewidth]{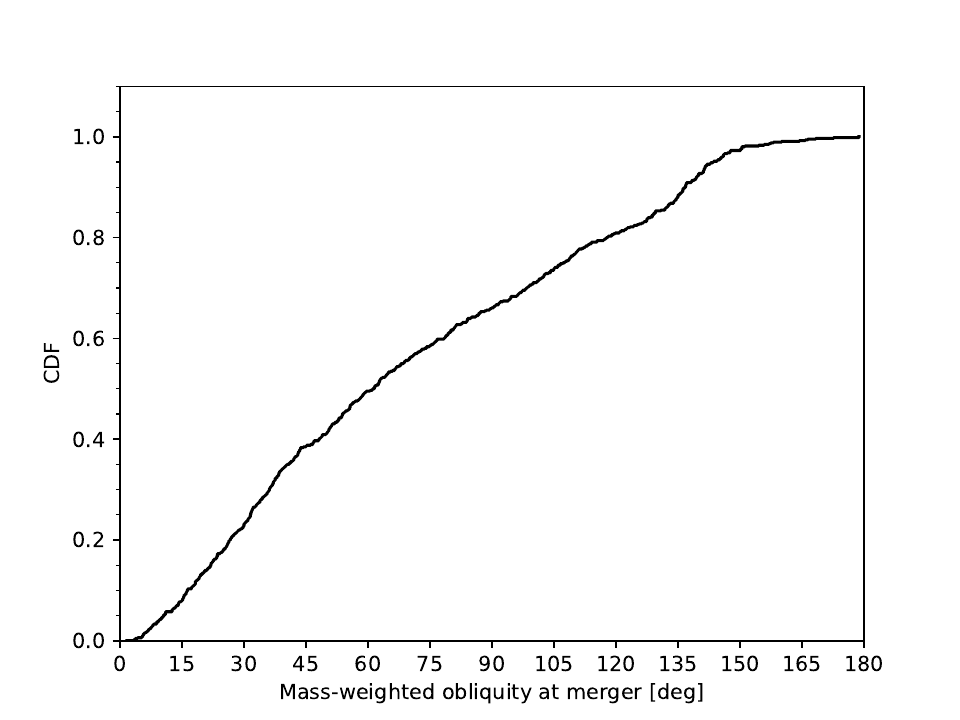}
    \caption{The cumulative distribution function of the angle between the orbital angular moment and the mass-weighted spin angle of all merging inner binaries. }
    \label{fig:triple-obliquity}
\end{figure}

Figure \ref{fig:triple-obliquity} shows the cumulative distribution
function of the obliquity (defined as the angle between the orbital
and mass-weighted spin angular momentum vectors) at the moment of
merger.  Misalignments are common, with the distribution spanning the
full range from aligned to polar to retrograde mergers. There is,
however, a modest preference for aligned configurations, with a median
obliquity of 60$^\circ$.

\subsection{Discussion} \label{subsec:triple_discussion}

\subsubsection{Merger delays}

Although a triple progenitor of \TOriC\, would plausibly have merged
within the required timescale, the merger times we obtain are in fact
shorter than our scenario would require. A merger after a few hundred
or thousand years after formation would effectively amount to a
scenario where no merger happened. Although mergers do happen at the
required lifetimes of a few 100 kyrs, especially in the intermediate,
rotation-influenced region, this requires a unfortunate amount of
fine-tuning.

A merger could be deferred during the pre-main sequence. The stars' larger radii and convective envelopes would increase tidal effects, effectively extending the stable region in Figure \ref{fig:triple-fate} towards larger semimajor axes, and moving inwards as they contract towards the ZAMS. 

At stellar masses of $\sim$16 M$_\odot$, the pre-main sequence lasts up to $\sim$0.3 Myr \citep[as obtained from {\sc sevn},][]{Iorio2023}, somewhat shorter than the $\gtrsim$0.5 Myr delay required in our scenario. 

Alternatively, the triple may exist in a (nearly) co-planar, vZLK-passive state for a relatively long period and then be perturbed into a vZLK-active state by a stellar encounter within the larger star cluster. Such an encounter could happen at any time, and does not require the merger to be delayed. 

In order to estimate the likelihood of such an encounter, we perform a
series of three-body scattering experiments \cite[using the {\sc
    scatter} package in {\sc
    starlab},][]{McMillan1996,1998A&A...337..363P} to determine the
cross section of interactions that can either lead to a direct merger
producing the present-day \TOriC, and interactions that can perturb
the orbit of \TOriC\, to an inclination susceptible to vZLK
oscillations. In the latter case, we assume that the inner binary of
the triple \TOriC\, is not perturbed during the encounter.

We use a semimajor axis of 17.59 au, an eccentricity of 0.6, and a mass of \TOriC2 of 12 M$_\odot$. For the collision experiment, both inner binary stars have a mass of 16.5 M$_\odot$, while in the perturbation experiment \TOriC1\, has a mass of 33 M$_\odot$ and the single star has a mass in the range of 1 to 35 M$_\odot$. We adopted the ONC environment to have a velocity dispersion of 3 km/s \cite[$\sim$3 km/s with a third dimension of similar magnitude,][]{Kim2019}. As collisional radii we use the ZAMS radii from {\sc SeBa}.

We find a cross section of $25\pi a^2$ for collision. For a perturbation to large inclination, we find a cross section of $83.95\pi a^2$ for perturbing stars of 10 M$_\odot$, which gradually drops to 0 for lower mass stars and $20\pi a^2$ for higher mass stars.

The collision-inducing encounter rate $\tau$ is then the product of cross section, the stellar density \cite[$\sim 2\cdot 10^4$ pc$^{-3}$,][]{Hillenbrand1998}, and the velocity dispersion. Together, this comes to a collision rate of $\sim$0.035 Myr$^{-1}$. For perturbations to large inclination, this comes to $\sim$0.12 Myr$^{-1}$. Although both are relatively rare over a 1 Myr period, the perturbation to large inclination is not implausible.

\subsubsection{Comparison to observed multiple population}

The proposed progenitor triple is not an extraordinary configuration;
the fraction of O-type stars with a companion in the decade around 1
au (approximately the range where we find mergers) is $\sim$0.3-0.4,
and the fraction of O-type stars in triples and higher order systems
is $\sim$0.7 \citep{Offner2022}.

The accelerated expansion scenario also prefers an equal-mass ratio
inner binary (which yields the largest stellar wind luminosity ratio,
as in Figure \ref{fig:luminosity_ratio}), which may be a natural
result of the formation of close binaries through fragmentation in a
disk \citep{Tokovinin2020}. On the other hand, observations from
\citet{Moe2017} show that there is only an excess twin
fraction\footnote{An excess twin fraction is here defined as an excess
in the mass ratio distribution $q>0.95$, compared to a (typically
decreasing) power law.} of both O and B stars for orbital periods
within 20 days. For our inner binary, that corresponds to a semimajor
axis of $\sim$0.5 au, which is exactly the range where we find that
circularization prevents mergers. Without an excess of twins the only
twins are the result of the high end of a power law
distribution. \citet{Moe2017} find that around a period of 100 days,
that power law gets close to that of a typical IMF distribution for O
stars (implying random pairing), but that it remains above that for B
stars. If our inner binary were a twin, the individual masses would
both be around the transition between O and B stars; regardless, the
remaining probability of twins is between 2 and 3\%. This does present
a challenge to our scenario being a probable solution to the proplyd
and shell lifetime problems; the preferred mass ratio of the inner
binary is most likely for vZLK-stable configurations, and uncommon in
vZLK-unstable configuration.

\subsubsection{Triple-induced mergers beyond the ONC}

Given that the circumstances that lead to our proposed vZLK-induced merger appear plausible and not specific to the ONC, it could happen in other star forming regions. As such an event can have substantial ramifications for the planet formation process inside the region through external photoevaporation and for other feedback-related processes, the intricate dynamics of stellar multiples can be important for the evolution of the surrounding star forming region and its stellar population. 

Similar triple (or other multiple) mergers are suggested for $\eta$ Carina, the site of the 1843 `Great Eruption' \citep{PortegiesZwart2016}, Source I in the nearby Orion Molecular Core 1 \citep{McCaughrean1997,Bally2017}, and V838 Mon \citep{Soker2003,Kaminski2021}.

The oblique magnetic field of \TOriC1\, is a result of the merger in
our model. Observations of O-type stars show that $\sim$10\% have
magnetic fields \citep{Grunhut2017}. However, as summarized by
\citet{Keszthelyi2023}, some magnetic stars are found in close
binaries, and the fraction of magnetic stars is constant with mass
whereas the fraction of multiples increases. This implies that the
number of magnetic stars formed through this channel is probably
small, and places an upper limit on the fraction of all O-type stars
formed this way of at most $\sim$1\%.

\subsubsection{Suggested present-day third component}

\TOriC\, has already been suggested to be a triple system by \citet{Vitrichenko2002} based on periodogram analysis, with a $\sim$1 M$_\odot$ companion in a 61.49 day orbit around \TOriC1. Its orbital separation of $\sim$0.8 au places it in the same region where our progenitor binary would have been. A confirmation of this companion would effectively falsify our model.

However, its period of 61.49 days coincides with a 4:1 resonance with the rotation period of \TOriC1\, \citep{Lehmann2010}, potentially making the signal a harmonic of the stellar rotation. A high-resolution observation using the {\sc gravity} instrument \citep{Gravity2018} failed to confirm its presence, although the angular resolution was comparable to the separation of $\TOriC1$ and the suggested companion.

Additionally, \citet{Lehmann2010} found a small inclination of the inner binary, making the relative orbit with \TOriC2 nearly polar \cite[the inclination of the orbit of \TOriC1\, and \TOriC2 is 98.9$^\circ$,][]{Balega2015}, making this system susceptible to the vZLK effect. We ran more triple simulations with the orbital parameters of \citet{Lehmann2010}. The initial conditions of these simulations can be found in Table \ref{tab:inner_ics}. Of 2000 runs, 1995 led to a merger with a median merger time of 135 years and a 99$^\text{th}$ percentile of 5,440 years. The remaining 5 simulations resulted in an unbound system within 60 years. Thus, we argue against the existence of the second, close companion to \TOriC1\, on the grounds that it would long ago have merged with the primary.

\begin{table*}
\begin{tabular}{|l|l|l|l|}
 & Inner binary & Outer binary & References \\
\hline
$M$ [M$_\odot$] & $\mathcal{N}\left(33.5,5.2\right)$, $\mathcal{N}\left(1.01,0.16\right)$ & $\mathcal{N}\left(12,3\right)$ & \cite{Lehmann2010,Balega2015} \\
$P$ [day] & $\mathcal{N}\left(61.49,0.02\right)$ & $\mathcal{N}\left(4011,112\right)$ & \cite{Lehmann2010} \\
$e$ & $\mathcal{N}\left(0.49,0.05\right)$ & $\mathcal{N}\left(0.61,0.08\right)$ & \cite{Lehmann2010} \\
$i$ [deg] & see text & $\mathcal{N}\left(99,2\right)$ & \cite{Lehmann2010} \\
$\Omega$ [deg] & $\mathcal{U}\left(0,360\right)$ & $\mathcal{N}\left(208.3,3.3\right)$ & \cite{Lehmann2010} \\
$\omega$ [deg] & $\mathcal{N}\left(198,6\right)$ & $\mathcal{N}\left(111,8\right)$ & \cite{Lehmann2010} \\
$M$ [rad] & $\mathcal{U}\left(0,2\pi\right)$ & $\mathcal{U}\left(0,2\pi\right)$ & \\
$P_{spin}$ & \multicolumn{2}{l|}{See text} & \\
$\hat{n}_{spin}$ & Aligned with inner binary & Aligned with outer binary \\
\end{tabular}
\caption{The initial conditions of the MCMC simulations of \TOriC\, with
  the proposed inner companion.}
\label{tab:inner_ics}
\end{table*}

\section{Conclusions} \label{sec:conclusion}

We have proposed a new scenario that can help explain both the proplyd lifetime problem in the ONC and the Orion Nebula shell lifetime problem. Our scenario involves the merger of a stellar binary into the present-day massive star \TOriC1, mediated by vZLK oscillations induced by the present-day binary companion \TOriC2. 

\begin{enumerate}
  \item The effective brightening of \TOriC1\, a few hundred thousand years ago can bring the expansion of the Orion Nebula's bubble into agreement with its stellar ages.
  \item A top hat model for the ISM density around the ONC can not reconcile the bubble's expansion with the stellar ages.
  \item The necessary brightening can not be fully explained by the merger of a progenitor binary, but a different degree of proplyd mass loading before and after the merger can increase the effective luminosity increase. We provide an estimate for the strength of this effect required to explain the brightening.
  \item The increase in ionizing luminosity, together with dynamical effects proposed by other authors, can resolve the proplyd lifetime problem; the slower expansion does not result in long shielding of the immediate neighourhood of \TOriC.
  \item vZLK-susceptible triples that would result in \TOriC\, on the merger of the inner binary result in the merger of the inner binary within 1 Myr in about half the configurations for a considerable range of inner semimajor axes.
  \item Mergers of such a \TOriC\, progenitor triple happen within a few kyr for a considerable part of the parameter space, but mergers can be delayed by increased pre-main sequence tidal effects, rotation, and perturbations by other cluster members.
  \item A vZLK-passive \TOriC\, progenitor triple would have a 10\% probability of being disturbed into a vZLK-active configuration by a cluster member.
  \item A solar mass inner companion of \TOriC1\, in a near polar orbit with \TOriC2, as has been proposed from spectra of \TOriC1, would rapidly merge with \TOriC1. We argue that this companion does not exist.
  \item Similar triple mergers have been conjectured for other massive stars, and the generality of the processes and conditions involved imply that similar scenarios can happen elsewhere. Considering the impact on star forming cloud disruption and protoplanetary disc evolution, this emphasizes the relevance of detailed orbital evolution on star and planet formation.
\end{enumerate}

\begin{acknowledgements}
We thank Thomas Haworth and Mordecai-Mark Mac Low and various
anonymous referees for their useful feedback on earlier incarnations
of this paper. We further thank Elko Gerville-Reache, Steven Rieder
(Anton Pannekoek Instituut), Lourens Veen (NLeSc), and Hanno Spreeuw
(NLeSc) their support in implementing {\sc tsumani} into the {\sc
  amuse} framework, and for updating it's buid system.

MJCW thanks Yorick Bleiji and Tim Fuchs for the coffee break that serendipitously originated this work, and Shuo Huang for his help in setting up the MCMC.

This work used the Dutch national e-infrastructure with the support of
the SURF Cooperative using grant no. EINF-10097, EINF-11368, and
EINF-13923, and by the Netherlands Onderzoekschool for Asteronomy
(NOVA) under project number 10.2.5.12.
Analysis was made using the open-source \texttt{Python} packages
\texttt{NumPy} \citep{2020Natur.585..357H} and \texttt{Matplotlib}
\citep{2007CSE.....9...90H}.

\end{acknowledgements}

\bibliographystyle{aa}
\bibliography{references}

\appendix

\section{Full sphere model posterior distributions} \label{sec:appendix1}

\subsection{Classic model}

\begin{figure*}
    \centering
    \includegraphics[width=0.66\linewidth]{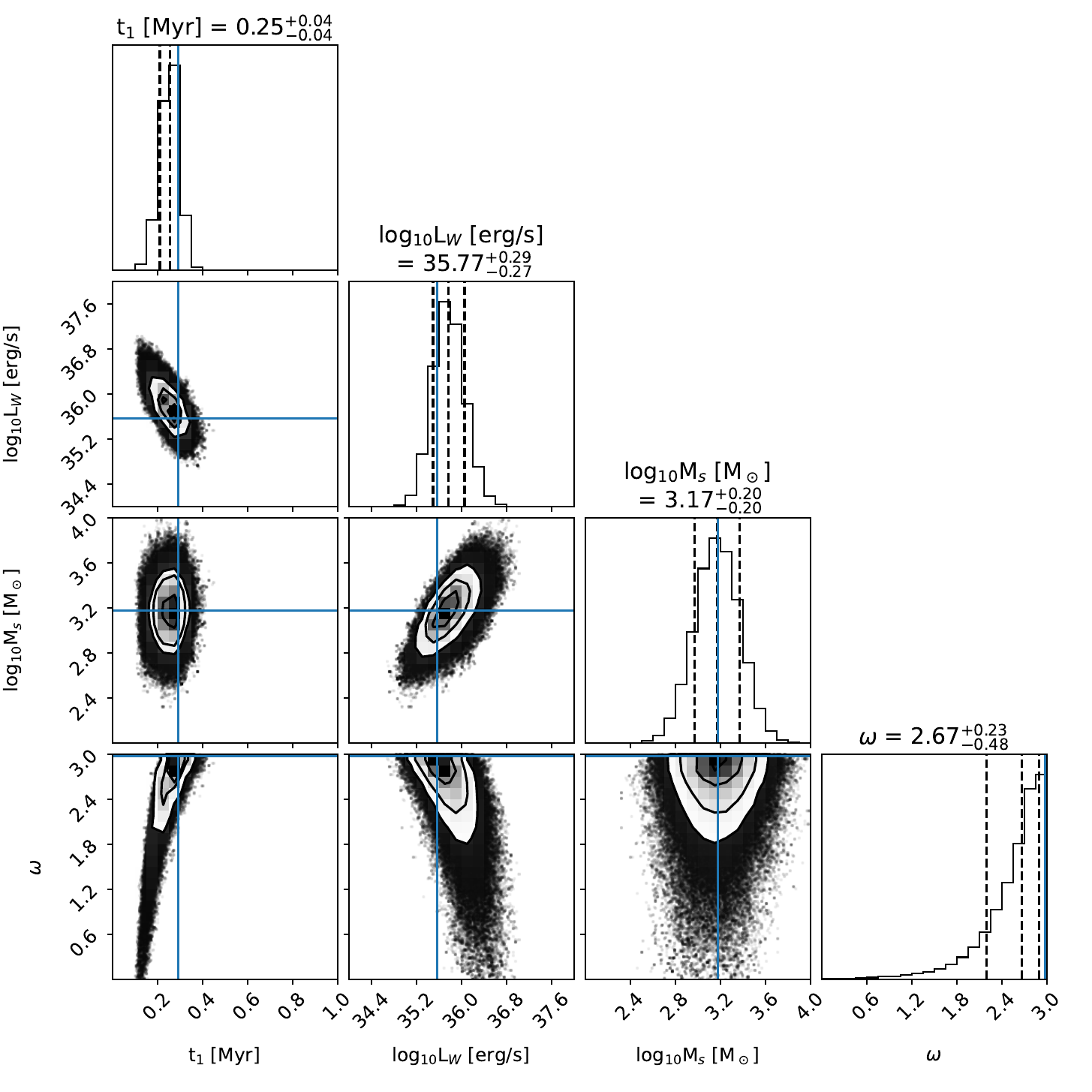}
    \caption{A corner plot of the posterior distribution of the classical full sphere model. Diagonal panels show one-dimensional probability distributions of each parameter, and off-diagonal panels show the two-dimensional probability distributions of the parameters in the same row and column. Blue lines indicate the parameters of the maximum-likelihood sample found by the MCMC.\\The parameters shown are the system age $t_1$, stellar wind luminosity $L_W$, the shell mass $M_s$, and the density power law $\omega$.}
    \label{fig:corner_classical_full}
\end{figure*}

In Figure \ref{fig:corner_classical_full} we show the posterior distribution of the classic full sphere model. As per usual, the wind luminosity and shell mass remain consistent. The system age is, as expected, constrained to smaller values than in the quarter sphere model, and large values are favoured even more for the density gradient.

\subsection{Brightening model}

\begin{figure*}
    \centering
    \includegraphics[width=\linewidth]{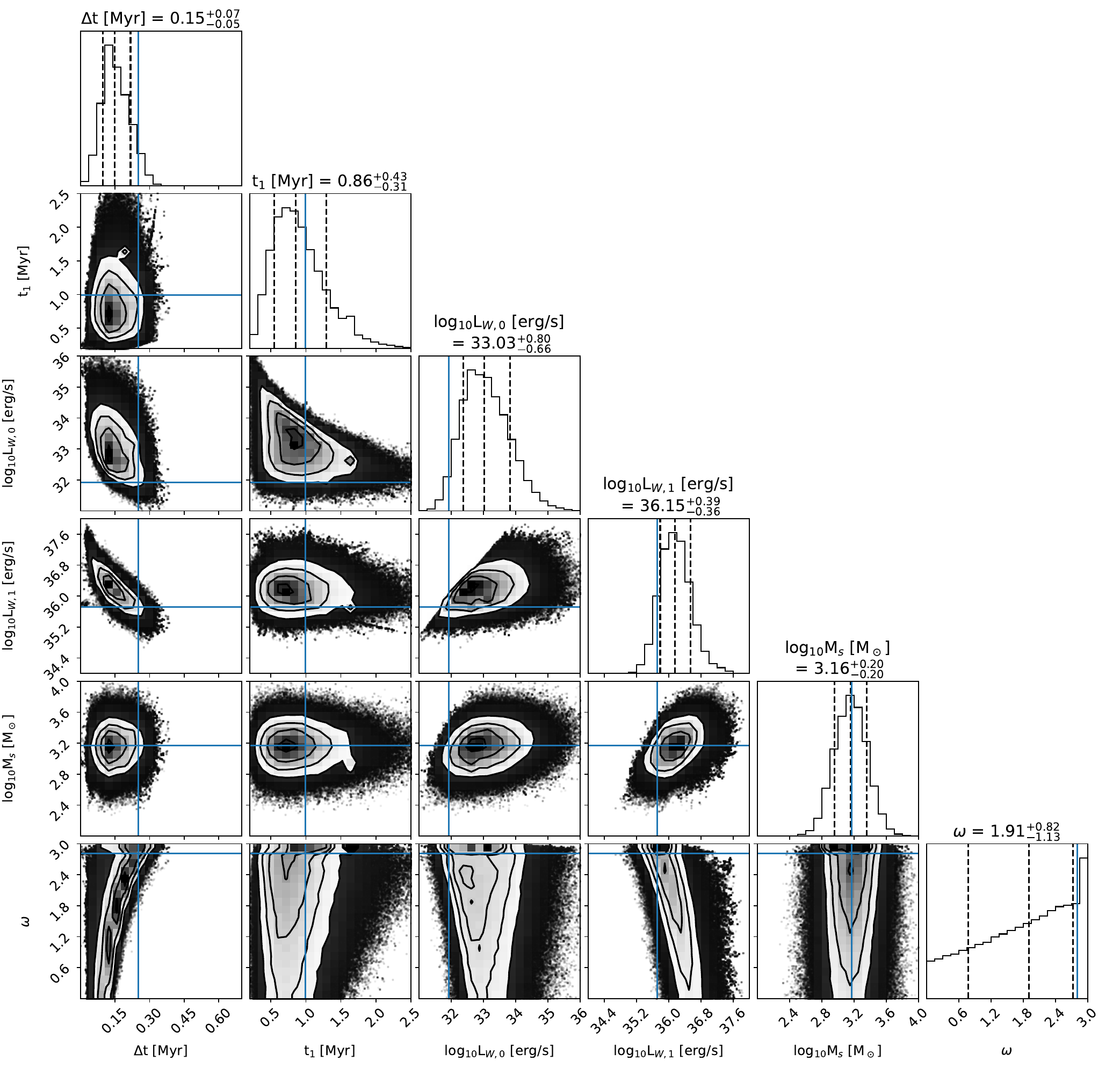}
    \caption{A corner plot of the posterior distribution of the brightening full sphere model. Diagonal panels show one-dimensional probability distributions of each parameter, and off-diagonal panels show the two-dimensional probability distributions of the parameters in the same row and column. Blue lines indicate the parameters of the maximum-likelihood sample found by the MCMC.\\The parameters shown are the time since brightening $\Delta t$, the system age $t_1$, the stellar wind luminosity before ($L_{W,0}$) and after ($L_{W,1}$) brightening, the shell mass $M_s$, and the density power law $\omega$.}
    \label{fig:corner_brightening_full}
\end{figure*}

In Figure \ref{fig:corner_brightening_full} we show the posterior distribution of the brightening full sphere model. As we are coming to expect, the wind luminosity and shell mass remain consistent. The time since brightening is slightly shorter than in the quarter sphere model. The system age tends to smaller values as well, though it is still broadly consistent with $\sim$1 Myr. Interestingly, the increase in stellar wind luminosity is more strongly constrained; here, the increase must have at least been by a factor of $\sim$1000.

\subsection{Top hat model}

\begin{figure*}
    \centering
    \includegraphics[width=\linewidth]{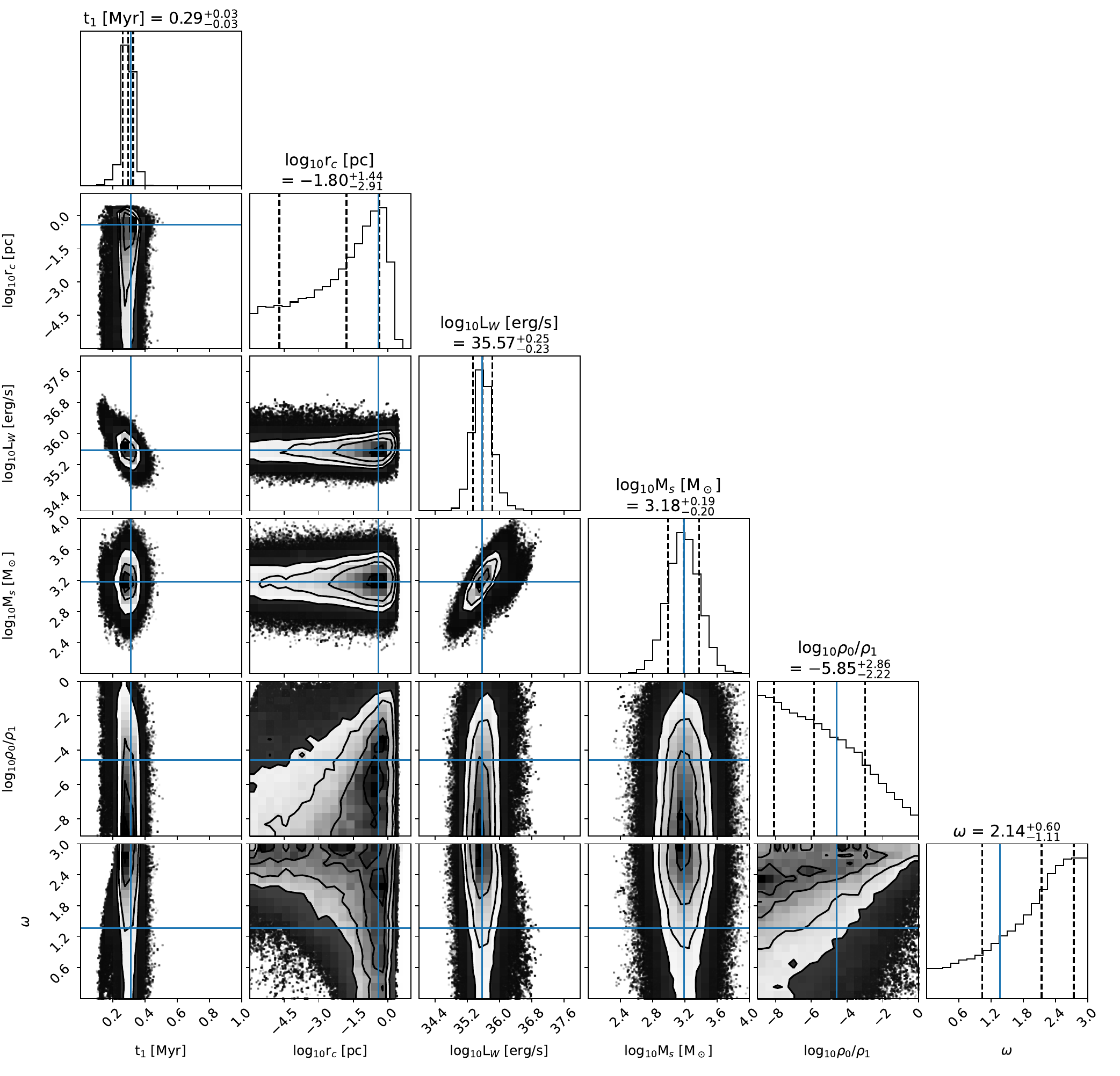}
    \caption{A corner plot of the posterior distribution of the top hat full sphere model. Diagonal panels show one-dimensional probability distributions of each parameter, and off-diagonal panels show the two-dimensional probability distributions of the parameters in the same row and column. Blue lines indicate the parameters of the maximum-likelihood sample found by the MCMC.\\The parameters shown are the system age $t_1$, the core radius $r_c$, the stellar wind luminosity $L_{W}$, the shell mass $M_s$, the density contrast between the core and beyond $\rho_0/\rho_1$, and the density power law $\omega$.}
    \label{fig:corner_tophat_full}
\end{figure*}

In Figure \ref{fig:corner_tophat_full} we show the posterior distribution of the top hat full sphere model. Completing the trend, the posteriors of the wind luminosity and shell mass remain consistent with the priors. The core radius, density ratio, and density gradient remain broadly unconstrained but with the same trends as in the quarter sphere model. As in the classic full sphere model, the system age is again constrained to small values.

\end{document}